\PassOptionsToPackage{table,dvipsnames}{xcolor}
 \documentclass[sigconf, table, natbib=true]{acmart}

\usepackage[utf8]{inputenc}
\makeatletter%
\@ifclassloaded{acmart}{
  \makeatletter
  \def\mdseries@tt{m}
  \makeatother
}{}
\makeatother

\usepackage[table,dvipsnames]{xcolor}
\usepackage{epsfig}
\usepackage{wrapfig}
\usepackage{pgfplotstable}
\usepackage{pgfplots}
\usepgfplotslibrary{groupplots}
\pgfplotsset{compat=newest} %
\usepackage{amsmath,mathtools}
\usepackage{amssymb}
\usepackage{relsize}
\usepackage{multicol}
\usepackage{bbm}
\usepackage[frozencache,cachedir=.]{minted}
\usepackage{float}
\usepackage{microtype}
\usepackage{tikz}
\usetikzlibrary{decorations.text,calc,shapes,arrows,arrows.meta, positioning,shapes.misc,decorations.markings,decorations.markings,decorations.pathreplacing,matrix,spy}

\usepackage{booktabs}
\usepackage{multirow}
\usepackage{adjustbox}
\usepackage{verbatim}
\usepackage[T1]{fontenc}
\usepackage[title]{appendix}
\usepackage{adjustbox} %
\usepackage{graphicx}
\usepackage{caption}
\usepackage{subcaption}
\usepackage[noend]{algpseudocode}
\usepackage{algorithm}
\usepackage{siunitx}
\usepackage{nicefrac}
\usepackage[colorinlistoftodos,disable]{todonotes}
\usepackage{spverbatim}
\usepackage{seqsplit}

\usepackage{array}
\newcolumntype{R}[2]{%
    >{\adjustbox{angle=#1,lap=\width-(#2)}\bgroup}%
    l%
    <{\egroup}%
}

\newcommand{\textcite}[1]{\citet{#1}}

\usepackage[autostyle, english=american]{csquotes}
\MakeOuterQuote{"}

\makeatletter
\pgfplotsset{
    every axis x label/.append style={
        alias=current axis xlabel
    },
    legend pos/outer south/.style={
        /pgfplots/legend style={
            at={%
                (%
                \@ifundefined{pgf@sh@ns@current axis xlabel}%
                {xticklabel cs:0.5}%
                {current axis xlabel.south}%
                )%
            },
            anchor=north
        }
    }
}
\makeatother

\newcolumntype{t}{>{\ttfamily}l}
\newcolumntype{T}{>{\ttfamily}c}

\newcolumntype{$}{>{\global\let\currentrowstyle\relax}}
\newcolumntype{^}{>{\currentrowstyle}}

\everypar{\looseness=-1} %
\newcommand{\bigO}{\ensuremath\mathcal{O}}

\hypersetup{draft}

\setcopyright{rightsretained}
\begin{document}

\acmDOI{}
\acmDOI{}
\acmISBN{}
\acmPrice{}
\acmConference[LEMINCS @ KDD'19]{KDD 2019 Workshop on Learning and Mining for Cybersecurity (LEMINCS'19)}{August 5th, 2019}{Anchorage, Alaska, United States}
\acmYear{2019}
\copyrightyear{2019}

\title{KiloGrams: Very Large N-Grams for Malware Classification}

  \author{Edward Raff}
  \orcid{0000-0002-9900-1972}
  \affiliation{%
    \institution{Laboratory for Physical Sciences}
  }
  \email{ edraff@lps.umd.edu}
  \affiliation{%
    \institution{Booz Allen Hamilton}
  }
  \email{ raff_edward@bah.com}
  
\author{William Fleming}
  \affiliation{%
    \institution{Laboratory for Physical Sciences}
  }
  \email{ william.fleming@lps.umd.edu}
    \affiliation{%
    \institution{U.S. Navy}
  }
  \email{ william.r.fleming1@navy.mil}
  
  \author{Richard Zak}
  \orcid{0000-0003-4272-2565}
  \affiliation{%
    \institution{Laboratory for Physical Sciences}
  }
  \email{ rzak@lps.umd.edu}
  \affiliation{%
    \institution{Booz Allen Hamilton}
  }
  \email{ zak_richard@bah.com}
  
  \author{Hyrum Anderson}
  \author{Bill Finlayson}
  \affiliation{%
    \institution{Endgame}
  }
  \email{hyrum@endgame.com}
  \email{bfinlayson@endgame.com}

    \author{Charles Nicholas}
  \affiliation{%
    \institution{Univ. of Maryland, Baltimore County}
  }
  \email{nicholas@umbc.edu}
  
  \author{Mark McLean}
  \affiliation{%
    \institution{Laboratory for Physical Sciences}
  }
  \email{ mrmclea@lps.umd.edu}
\renewcommand{\shortauthors}{E. Raff et. al.}

\begin{abstract}
N-grams have been a common tool for information retrieval and machine learning applications for decades. In nearly all previous works, only a few values of $n$ are tested, with $n > 6$ being exceedingly rare. Larger values of $n$ are not tested due to computational burden or the fear of overfitting. In this work, we present a method to find the top-$k$ most frequent $n$-grams that is 60$\times$ faster for small $n$, and can tackle large $n\geq1024$. Despite the unprecedented size of $n$ considered, we show how these features still have predictive ability for malware classification tasks. More important, large $n$-grams provide benefits in producing features that are interpretable by malware analysis, and can be used to create general purpose signatures compatible with industry standard tools like Yara.  Furthermore, the counts of common $n$-grams in a file may be added as features to publicly available human-engineered features that rival efficacy of professionally-developed features when used to train gradient-boosted decision tree models on the EMBER dataset.
\end{abstract}

\maketitle

\section{Introduction}

In this work, we are interested in the task of finding the top-$k$ most frequent $n$-grams in a large corpus. Given a corpus $\mathcal{C}$ of documents, and an alphabet $A$, there are $|A|^n$ possible $n$-grams, making the use of large $n >6$ computationally infeasible for many applications. Still, $n$-grams have been a bread-and-butter tool for natural language processing and other related fields for decades, thanks to their simplicity and usefulness. As such, significant work has gone into engineering systems to work with $n$-grams \cite{JoDI22,Buck2014,Pauls:2011:FSN:2002472.2002506}. This is also true for malware classification, where we wish to determine whether a file is benign or malicious (malware detection), or to identify the specific family of a known malicious file (malware family classification). 

In particular, we are interested in selecting $n$-grams for large values of $n$. This is motivated by the use of byte $n$-grams as features for malware classification. There has long existed an intuitive need for larger values of $n$ in this space due to the nature of content encoded in executable file formats. For example, if we consider byte $n$-grams for Microsoft Windows Portable Executable (PE) files, one x86 assembly code instruction could be up to 15 bytes long. This would require us to consider at least 16-grams to capture this \textit{one} instruction in context. Early work determined large values like $n=15$ performed best \cite{Abou-Assaleh2004a}, but this was only possible because of the small corpus size (36.9 MB). \textcite{Goldberg1998} proposed using 20-grams since the average malware detection signature used in 1998 was 20 bytes in length. The seminal BitShred clustering work proposed 16-byte grams, but needed a cluster of 64 machines to scale past 60,000 files, and the use of feature hashing\cite{Weinberger2009a} meant they did not have the original features \cite{Jang2011}. As the size of malware corpora has grown, the exponential cost in increasing the value of $n$ has forced researchers to consider small values of $n$ and other alternatives. Recent works that have looked at corpora with at least 400,000 files have been constrained to 6-grams or less \cite{raff_ngram_2016}. Considering that the Anti-Virus (AV) industry is making use of datasets that range in size from ten million \cite{Li2017} to hundreds of millions \cite{tagkey2014iv} of files, the methods that exist today simply can't scale to the magnitude of industry corpora, and old results using hundreds of files are not sufficient to base decisions on. 

In this work, we introduce the KiloGram technique for efficiently finding the top-$k$ most frequent $n$-grams for large values of $k$ and $n$ with high probability under the assumption of a power-law distribution to the $n$-grams. If $L$ is the total number of observed $n$-grams, or bytes, in the corpus, our algorithm will take only $\bigO(L)$ time and $\bigO(B + k \cdot n)$ memory. The parameter $B$ is a budget factor to control the accuracy of the method, and since $n \ll k \ll B \ll L$, this memory 
cost is minimal. For our tests, for example, this $B$ corresponds to using $\approx$9 GB of RAM to extract frequent $n$-grams from 5 TB of data.  For $n\in [2, 8]$, our approach is 60 times faster than previous works, and runtime does not increase with $n$, allowing us to test $n=1024$ and beyond.  
This allows us to answer questions about the behavior of byte based $n$-grams in a more conclusive way than prior work \cite{Zak2017}. 

In \autoref{sec:related_work}, we review related works that aim to increase the value of $n$ for malware classification. The proposed KiloGram algorithm will be presented in \autoref{sec:kilograms}. We perform the first ever investigation of large $n \in [8, 1024]$ for malware classification in \autoref{sec:results}. We found that $n=8$ performs well and generalizes well over three years of concept drift in malware. Ourr results show that the common assumption that $n$ should be larger in the malware analysis space do not hold for larger modern corpora. Surprisingly, $n=1024$ also results in nontrivial malware classification accuracy. We demonstrate in \autoref{sec:analysis} that large $n$-grams are interpretable to malware analysts, can help them automate laborious and error prone parts of their job, and can be combined with lower-effort domain knowledge to rival (as measured using the EMBER dataset) proprietary industry feature extractors built from decades of expertise. Finally we will conclude in \autoref{sec:conclusion}.

\section{Related Work}\label{sec:related_work}

N-grams have been used as features for malware analysis since the first work in automating malware detection in 1995 \cite{Kephart:1995:BID:1625855.1625983}, and have consistently been used for malware classification systems ever since \cite{Reddy2006,Kolter:2006:LDC:1248547.1248646,Tahan:2012:MAM:2188385.2343677,raff_ngram_2016}. Except when using small datasets (hundreds of MB or less), values of $n>8$ are never tested in published literature due to their computational burden. However an intrinsic concern is that $n$-grams need to be larger. For example \citet{5645851} noted that a byte 6-gram was too short to fully capture an observed x86 instruction 2.4\% of the time. 

In attempts to increase the $n$-gram length, some have developed techniques that attempt to coalesce multiple $n$-grams to a single canonical base form. One example of this is $n$-perms \cite{Karim2005}, where a sorted ordering is applied to every $n$-gram to map them to a single canonical form (e.g., \textit{ACB}, \textit{BCA}, and \textit{CAB} would all map to \textit{ABC}). This $n$-perm approach has been used for malware classification with a value of $n$ as large as 10 \cite{walenstein2007exploiting}.  

While our focus is on processing the raw bytes of a binary, $n$-grams have been popular for assembly instructions as well. 
Similar coalescing techniques have been necessary to do any work with assembly due to computational constraints. Prior works have examined replacing all memory addresses, register references, and constants with generic \texttt{mem}, \texttt{reg}, and \texttt{const} placeholders \cite{Masud2008}, though it is more common to remove all instruction operands entirely 
\cite{Shabtai2012}.

While much work has gone into storing and processing known $n$-grams efficiently \cite{JoDI22}, little has been been done to try to extend the value of $n$ itself in a time and memory efficient manner. The only prior work we are aware of 
was performed by \citet{Nagao1994}. They considered obtaining $n \in [1, 255]$ by cleverly converting the most frequent $n$-gram calculation into a sorting problem, resulting in $\bigO(L \log L)$ complexity and $\bigO(L)$ space. While exact, these bounds are worse than ours and necessitate slower out-of-core sorting than the proposed method.  Furthermore, the method is limited to $n\leq 255$, and only tested up to $n=10$. 
Our method relies on certain distributional assumptions to hold with high probability, but allows us significant speed and practicality benefits with $\bigO(L)$ time and $\bigO(B + k \cdot n)$ memory. 

\section{KiloGramming } \label{sec:kilograms}

Our goal is to find the top-$k$ most frequent $n$-grams for large values of $n$. To do this, we build off of two prior works. 
First is the hash-gram approach \cite{hashgram_2018}. Hash-grams find the top-$k$ most frequent \textit{hashes} of $n$-grams. They created a large table of size $B=2^{31}-19$ to store hashes, and simply ignored collisions. By using a rolling hash function $h(\cdot)$ \cite{Cohen:1997:RHF:256163.256168} , they were able to obtain orders-of-magnitude speedup over normal $n$-gram tabulation, at the cost of losing information about what $n$-grams are actually being used. 

The hash-gram approach works under the common assumption that $n$-grams follow a Zipfian (power law) distribution \cite{Zipf1949}. The Zipfian distribution has probability mass function $f(\cdot)$ and cumulative distribution function $F(\cdot)$ given by
\begin{multicols}{2}
\noindent
\begin{equation}
    f(x; p, |A|)  = \frac{x^{-p-1}}{H_{|A|}^{(p+1)}}  \label{eq:pdf_zipf}
\end{equation} 
\begin{equation}
    F(x; p, |A|) = \frac{H_{ x }^{(p+1)}}{H_{|A|}^{(p+1)}} \label{eq:cdf_zipf}
\end{equation}
\end{multicols}
\noindent where $H_z^{(p)} = \sum_{i=1}^z i^{-p}$ indicates the $z$'th harmonic number of the $p$'th order, and $x \in [1, 2, \ldots, |A|]$.

Under the Zipfian-distributed assumption, it was shown that hash-grams discover the correct top-$k$ hashes with high probability \cite{hashgram_2018}. The Zipfian distribution is a surprisingly good fit to human language and many other tasks \cite{Piantadosi2014}, and as such has been a common and useful model for $n$-gram based features in natural language processing\cite{Cavnar1994}, as well as for $n$-grams over bytes from binary executables \cite{raff_ngram_2016}. 

Naively, one would like to use an approach such as the Space-Saving algorithm \cite{Metwally:2005:ECF:2131560.2131596}, which can return the top-$k$ most frequent items from a stream.
At a high level, it works as a kind of 'rank' based cache. If an item is in the Space-Saving data structure, its rank is increased as well as an associated count. If an item is not in the cache, the current item with the lowest rank is replaced, its rank increased, and it's error bound reset. Based on the current error bounds, it can estimate top-$k$ most frequent items in a stream, and in some cases guarantee that they are the true top-$k$. Thanks to clever design, updates to the Space-Saving data structure are $\bigO(1)$. 
In this scenario, one would treat all possible $n$-grams as the stream to process, and select the top-$k$ after processing the stream\footnote{Small scale tests on 80,000 files found that the computational overhead of the Space-Saving structure is also significant. An attempt to find the top $k=10,000$ and $n=6$ with $B=1,000,000$ in this scenario took just as long as computing the exact $n$-grams in the first place, and failed to return any of the true top-$k$ due to difficulty is knowing the correct budget size since the $\bigO$ notation hides constant factors. }. 
However this becomes computationally intractable as $k$ increases, and for a Zipfian distribution with $p=0$, the Space-Saving algorithm requires $B=\bigO \left( k ^ { 2 } \log ( | A | ) \right)$ buckets to obtain the true top-$k$ $n$-grams, resulting in $\bigO \left(n k ^ { 2 } \log ( | A | )  \right)$ memory use. 
When we consider that the size of our alphabet is a function of the $n$-gram size (i.e., $|A|=256^n$), we get $B=\bigO  \left(  n k ^ { 2 } \right)$ and a total memory use of $\bigO \left(n^2 k ^ { 2 }  \right)$, which is not tenable if we wish to consider larger $k$ or large $n$, let alone both as we do in this work. 
Prior works have used $k=8,000$ as the largest $k$ \cite{Cafaro2018}, which is insufficient for feature selection of $n$-grams where we need to preserve  $k\geq100,000$.

To resolve these issues, we introduce the \textit{KiloGram} algorithm. This algorithm enables $n$-gram computation with $n$ exceeding 1000 by extending the hash-gram approach with a second pass that selectively leverages the Space-Saving algorithm.  Its run-time complexity is $\bigO(L)$, with two iterations over the corpus to process $n$-grams and place them into a hash-table (first pass) or Space-Saving data structure (second pass): either insertion is $\bigO(1)$ complexity. The other operations in the proposed method are $\bigO(B)$ (e.g., quick-select), and since $B< L$, we arrive at $\bigO(L)$ total complexity. For memory, we require $\bigO(B)$ memory for the large table $T$, and an additional $\bigO(k \cdot n)$ memory for the storage of exact $n$-grams in the space-saving data structure, so that memory complexity is  $\bigO(B+ k \cdot n)$. 

\begin{algorithm}[!ht]
\caption{KiloGramming}
\label{algo:kilo_gram}
\begin{algorithmic}[1]
\Require Bucket size $B$, rolling hash function $h(\cdot)$,  corpus of  $\mathcal{C}$ documents, and desired number of frequent hash-grams $k$, and hashing stride $s$.

\State T $\gets $ new integer array of size $B$
\For{ all documents $x \in \mathcal{C}$} \Comment{$\bigO(L)$ for $L$ total $n$-grams}
	\For{$n$-gram $g \in x$} 
		\State $q' \gets h(g) \mod B$
		\If{$q' \mod s = 0$} \Comment{Hashing-Stride check}
		    \State $T[q'] \gets T[q'] + 1$ 
		\EndIf
    \EndFor
\EndFor
\State $T_k \gets \text{QuickSelect}(T, k)$ %
\State $S \gets $ new Space Saving structure with $B_S$ buckets. 
\For{ all documents $x \in \mathcal{C}$} \Comment{Second pass over data}
	\For{$n$-gram $g \in x$} 
		\State $q' \gets h(g) \mod B$
    	\If{$q' \in T_k$}
    	    \State Insert $g$ into $S$
    	\EndIf
    \EndFor
\EndFor
\State \Return top-$k$ entries from S
\end{algorithmic}
\end{algorithm}  

The pseudo-code is given in \autoref{algo:kilo_gram}. On the first pass through the dataset, we use the hash-gram approach of creating a large table to find the top-$k$ most frequent hashes, which under the assumptions of a Zipfian distribution, will find the true top-$k$ hashes with high probability \cite{hashgram_2018}. The hash-graming corresponds to lines 1--4 and line 6. Line 5 is an addition we will discuss soon in \autoref{sec:hash_stride}. 

Once we have the set of the top-$k$ hashes, we create a new Space-Saving data structure to help us keep track of the corresponding top-$k$ $n$-grams. We will perform a second pass over the data, and use the top-$k$ list of hashes as a white list for the Space-Saving algorithm. In this way the majority of observed $n$-grams will not be processed because they do not have one of the specified hash values, and the Space-Saving structure allows us to filter out the collisions from the true most-frequent $n$-grams. We require only $\bigO(k)$ buckets in the Space Saving structure for all practical use cases, which we prove in \autoref{sec:math}, resulting in $\bigO(k \cdot n)$ memory use for the second step. This dramatically reduces the amount of memory required, and runs orders of magnitude faster than attempting to use the Space-Saving approach on the entire corpus. 
The second pass over the data requires less time to run than the first pass because fewer memory accesses are being performed ($\geq$99.99\% of $n$-grams are non-frequent \cite{raff_ngram_2016}), and these memory accesses result in more cache hits (smaller Space-Saving structure compared to large array $T$). In testing, the second pass can account for as little as 9.76\% of the total runtime. 

\subsection{Hashing-Stride} \label{sec:hash_stride}
We introduce the concept of a \textit{hashing-stride} of size $s$ to further enhance the utility of the $n$-grams found so that they are useful for creating features. The application of the hash-stride is simple. For each $n$-gram $g$, we will compute its hash $q = h(g)$. If $q \mod s \neq 0$, the $n$-gram is discarded.
Thus, hash-striding is simply a deterministic downsampling of input $n$-grams by a factor of $s$.  

Hash-striding is important to reduce  redundancy caused from the sliding window effect across long common sequences.  In particular, for a ubiquitous sequence of length $\ell > n$, the resulting top $k$ $n$-grams would be dominated by $\ell-n+1$ equally frequent and essentially redundant sub-sequences.  Including these $n$-grams in the top $k$ effectively reduces $k$ by a factor of $(\ell-n)$.

A naive alternative to reduce the number of $n$-grams considered for the top-$k$ is to use a spatial stride $z$, where one steps by a constant number of $z$ grams through the input sequence. However, if a frequent $n$-gram does not occur at intervals of exactly $z$, this approach would fail to identify occurrences of the $n$-gram, resulting in inaccurate counts or in the worst case, exclusion.  By using a hashing-stride of $s$, we reduce the total expected number of unique $n$-grams to process by a factor of $s$. This is because for any particular $n$-gram $g$, we will always count its occurrence regardless of its offset within a file. This ensures that counts of $n$-grams are accurate. %

From an implementation perspective, hashing-stride allows one to perform a necessary first approximation to feature selection without having to perform any kind of communication or coordination between files, and without any additional significant computation.  This also means we are technically selecting the top-$k$ $n$-grams from $|A'|/s$ unique $n$-grams, where $A'$ is the set of observed $n$-grams from the possible alphabet $A$ (i.e., $|A'| \leq \min(L, |A|)$). We will continue to refer to this as just the "top-$k$" for brevity. For all experiments, unless stated otherwise, we use a hash-stride of $s=\lceil n/4 \rceil$.

\subsection{KiloGrams under the Zipfian Distribution} \label{sec:math}

We now prove that
\autoref{algo:kilo_gram} preserves the correct top-$k$ $n$-grams when $A$ follows a Zipfian distribution.  In what follows, $L \ge |A'|$ represents the total number (including duplicates) of $n$-grams in the corpus.
In the proof (see \autoref{sec:kil_bound_proof}), it is 
assumed that the first pass of the algorithm has obtained the true top-$k$ hashes of the top-$k$ $n$-grams, which was previously proven to occur with a high probability \cite{hashgram_2018}. The proof continues by showing that given the true top-$k$ hashes and $p\geq 1$, the expected number of colliding non-frequent $n$-grams (including duplicates) is upper bounded by
\begin{equation} \label{eq:final_bound}
    6 L / \left( B \pi^2\right).
\end{equation}
Thus we may preserve the true top-$k$ by having a sufficiently large hash-table to disambiguate the frequent and non-frequent collisions. 

Since our implementation is in Java we  use $B=2^{31}-19$, the largest prime array size allowed by Java. This value is also realistic and requires only \textit{8.6 GB} of RAM, well within the capacity of a modern laptop. With this, \autoref{algo:kilo_gram} across one \textit{petabyte} of $n$-grams  ($L=10^{15}$), we would expect at most 283,100 collisions. As such, adding a constant of 300,000 to the size of the Space-Saving structure $S$ in \autoref{algo:kilo_gram} should suffice for any application which could practically run on a single computer. We include  $3\cdot k$ as an alternative hedge against any situation where our empirical data does not follow a power-law type distribution. Thus, in experiments, we use $B_S=\textrm{max}(k + 300000, 3 \cdot k)$ buckets. 
In all of our experiments, the bound shown in \autoref{eq:final_bound} was never violated.

The Space-Saving structure is unnecessary for the proof, but included to ensure our approach will work should the true distribution depart from a Zipfian distribution. 
The Space-Saving algorithm also allows for a similar bound on the total number of buckets needed. Following the proofs in \cite{Metwally:2005:ECF:2131560.2131596}, and noting that $B>n \cdot k$ for all of our experiments, we  reach a bound that $B_S = \bigO(k)$ for all possible Zipfian data streams. 

\subsubsection{Derivation of KiloGram Bound} \label{sec:kil_bound_proof}

For a uniform hash function, the expected number of collisions for any individual bucket is $L/B$. There are $k$ buckets of interest corresponding to the top-$k$ most frequent $n$-grams. 
If $f(x; p ,|A|)$ is the Zipfian probability distribution function with cumulative distribution $F(\cdot)$, the total number of observed $n$-grams that do not collide with the top-$k$ is $ L \cdot (1-F(k; p, |A|))$. Then, the expected number of infrequent $n$-grams that collide with the top-$k$ $n$-grams is then equal to this value times $k/B$:

\begin{equation} \label{eq:expected_collisions}
    k \cdot L \cdot \left(1-\frac{H_k^{(p+1)}}{H_{|A|}^{(p+1)}}\right)/B
\end{equation}

\autoref{eq:expected_collisions} includes multiple occurrences of the same infrequent $n$-gram, and so is pessimistic. If one makes the Space-Saving algorithm large enough to handle all possible collisions, then the true top-$k$ $n$-grams are obtained with high probability. This is because the Space-Saving algorithm degrades to a simple hash-table that counts everything exactly when the number of buckets in the Space-Saving data structure is greater than or equal to the number of unique items in the table. 

From a theoretical perspective, using the Space-Saving algorithm instead of a hash-table gives us added flexibility to deal with the rare possibility of having more than the expected number of $n$-gram collisions. More practically, we use the Space-Saving algorithm because real data is not \textit{truly} Zipfian distributed, and this gives us a method of gracefully handling deviations from theoretical expectations. 

Considering the expected collisions in \autoref{eq:expected_collisions}, we can make some practical simplifications given hardware constraints and data assumptions. First, we pessimistically assume that $p=1$, which is the worst case for "interesting" power law distributions observed in real data, which generally fall in the range of $[1, 4]$ \cite{doi:10.1137/070710111}. 

Next, we pessimistically assume that the alphabet $A$ is infinite in size. This technically degrades to the Zeta distribution, and is pessimistic because it maximizes the amount of probability mass that exists in the tail of the distribution (i.e., reduces the value of $F(k;, p, |A|)$). Doing so, we obtain
$$
\underset{|A|\to \infty }{\text{lim}}- k \cdot L \cdot  \left(\frac{H_k^{(2)}}{H_{|A|}^{(2)}}-1\right)/B = \frac{k L \left(\pi ^2-6 H_k^{(2)}\right)}{B \pi^2}
$$
\noindent which further simplifies to 

$$
\frac{6 \cdot k \cdot L \cdot \psi ^{(1)}(k+1)}{B \pi ^2}
$$

\noindent where $\psi ^{(\alpha)}(\beta)$ is the PolyGamma function. This simplification is significant, because $\forall k \geq 1$, $k^{-1} > \psi ^{(1)}(k+1)$, allowing us to replace the PolyGamma function with a pessimistic upper bound. Further, because  $\underset{k\to \infty }{\text{lim}}\left(\frac{1}{k}/\psi ^{(1)}(k+1)\right) = 1$, this upper bound is tight.
We may further simplify by replacing the PolyGamma evaluation with $1/k$, yielding \autoref{eq:final_bound}.

The bound in \eqref{eq:final_bound} states that the number of collisions is linear in the total number of $n$-grams processed. This is not a surprising result.  More important is that we have a numerical upper bound that can be employed to reduce the number of collisions dramatically.

Under a more pessimistic assumption that $p< 1$, the \eqref{eq:final_bound} bound would not hold. However, a similar result can be obtained using the Spave-Saving structure. For that algorithm, Zipfian data with $p=0$ (the worst possible case) requires $\bigO(k^2 \log(|A|))$ buckets. Since we have already found the top-$k$ colliding hashes based on a table of size $B$, plugging this into the proof from \cite{Metwally:2005:ECF:2131560.2131596} leads to a requirement of $\bigO(B^{-1} k^2 \log(|A|))$ buckets. Noting that for processing $n$-grams, $|A| = 256^n$, this can be simplified to $\bigO(B^{-1} n k^2)$. We note that in all experiments in this paper, $B > n \cdot k$, which allows those terms to cancel. This leaves the Kilogram approach with a total of $\bigO(k)$ buckets to process any Zipfian dataset for all $p\ge0$, a considerable improvement compared to $\bigO(n k^2)$ buckets for the Space-Saving algorithm alone.

\section{Classification Results} \label{sec:results}

To test and evaluate the proposed KiloGram approach, we  make use of four datasets that include Windows PE files and Adobe Portable Document Format (PDF) files. The datasets  are summarized in \autoref{tbl:dataset_summary}, with more detail in the appendix. 

\begin{table}[ht]
\caption{All datasets used in our experiments, including size of training and testing sets, and primary year the data is from. }
\label{tbl:dataset_summary}
\begin{adjustbox}{max size={1.00\columnwidth}{0.85\textheight}}
\begin{tabular}{@{}lcrrr@{}}
\toprule
Dataset      & \multicolumn{1}{c}{Year} & \multicolumn{1}{c}{Train} & \multicolumn{1}{c}{Test} & Storage Size\\ \midrule
Industry EXE   & 2014-2015 & 2,011,786                 & 400,000      & 5 TB            \\
EMBER          & 2017 & 600,000                   & 200,000         & 936 GB         \\
Public PDF     & 2018 & 75,1829                   & 83,780            & 464 GB       \\ 
VirusShare-20C & 2013-2018 & 160,000                   & 40,000       & 141 GB              \\ \bottomrule
\end{tabular}
\end{adjustbox}
\end{table}

The "Industry EXE" dataset was provided to us, under a non-disclosure agreement, by a third party AV company. The training set contains 2 million Windows PE executables, evenly split between benign and malicious \cite{raff_shwel}, and a test-set of 400,000 binaries, also evenly split\cite{raff_ngram_2016}.

The files from which the EMBER dataset \cite{Anderson2018} were created can be obtained from VirusTotal \cite{Virustotal}. 
EMBER has an even split between benign and malicious, and since it is 2-3 years newer than Industry EXE, we can use it as an extreme test of generalization over time. This is important since malware is known to exhibit concept drift \cite{203684}.  

Our "Public PDF" dataset was constructed from VirusShare for PDF malware  \cite{VirusShare} (19\% of the data) and using Common Crawl\footnote{\url{http://commoncrawl.org}} for benign PDF files. 

Our "VirusShare-20C" (or "VS-20C") dataset was constructed from VirusShare \cite{VirusShare}, using AVclass to identify 20 PE malware families with exactly 10,000 total samples each \cite{Sebastian2016,Seymour2016}.

We use all four datasets in our evaluations throughout the paper, and observe consistent results across each in terms of the nature of larger $n$-gram sizes. For clarity, we consider each dataset in turn to highlight  results and behavior across the four sets. Following \cite{raff_ngram_2016}, we use Elastic-Net regularized logistic regression \cite{Zou2005,Yuan2012,JMLR:v18:16-131} to train predictive models from the byte $n$-grams. Using the elastic net regularizer of $\|w\|_1 + 0.5 \|w\|_2^2$ provides an important feature selection as part of the model training process, as the $\|w\|_1$ term will shrink insignificant features to  $0$, and provides empirical and theoretical robustness to high-dimensional problems with noisy and irrelevant features \cite{Ng2004}. To parllelize the Kilo-Gram algorithm, we use the approach specified in \cite{raff_hash_gram_parallel} for lines 1-6, and naive parallelization of the Space-Saving algorithm using the approach of \cite{Cafaro2018} to merge Space-Saving data structures for lines 8-13. QuickSelect is run as a single thread. 

For all datasets we use \textit{balanced accuracy} \cite{Brodersen:2010:BAP:1904935.1905533}, which re-weights the test data as if there were an equal number of files in all test sets. This is done to make the accuracy number comparisons more meaningful across each dataset, where there may be slightly different ratios of benign-to-malicious files. For our binary classification problems, we will also use the Area Under the ROC Curve (AUC) \cite{Bradley1997}. This metric is of particular interest in malware detection, since one wishes to select a threshold that corresponds to low false positive rates, and AUC is the integral of true positive rate across all false positive rates, without requiring one to select a threshold \textit{a priori} (in contrast to accuracy).

\subsection{Hashing-Stride Improves  Performance}

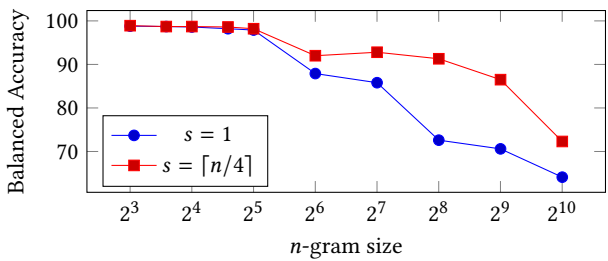
\begin{figure}[!h]
\centering
\begin{tikzpicture}
\begin{axis}[
    xlabel=$n$-gram size,
    ylabel=Balanced Accuracy,
    legend pos=south west,
    xmode=log,
    log basis x={2},
    height=4cm,
    width=\columnwidth,
    ]
    
    \addplot+[] coordinates {
        (8,98.8)
        (12,98.7)
        (16,98.6)
        (24,98.2)
        (32,97.9)
        (64,87.9)
        (128,85.8)
        (256,72.6)
        (512,70.6)
        (1024,64.1)
        };
    
        \addplot+[] coordinates {
        (8,98.9)
        (12,98.7)
        (16,98.7)
        (24,98.6)
        (32,98.2)
        (64,92)
        (128,92.8)
        (256,91.3)
        (512,86.5)
        (1024,72.3)
        };

    \legend{$s=1$, $s=\lceil n/4 \rceil$}
    
\end{axis}
\end{tikzpicture}
\caption{Balanced Accuracy results (y-axis) on the Public PDF dataset as we increase the $n$-gram size (x-axis, log-scale), and alter the hashing stride $s$. Using a hashing-stride retains more performance as $n$ becomes larger. }
\label{fig:Hash_stride pdf}
\end{figure}

First, we evaluate the inclusion of our hashing-stride approach, as discussed in \autoref{sec:hash_stride}. The expectation is that, as $n$ becomes larger, the performance of models built from the top-$k$ most frequent features will drop due to an increasing redundancy in the top-$k$ list. Our results back up this theoretical prediction, as shown in \autoref{fig:Hash_stride pdf}. 

\autoref{fig:Hash_stride pdf} shows that for  small $n \leq 16$, the absolute difference in accuracy is less than 0.1 in all cases, and the hashing-strides are correspondingly small values $s \in [2, 4]$. At $n=32$ the performance gap  increase slightly, and by $n=64$ the difference becomes significant. Across all $n \in [8, 1024]$, the use of a hashing-stride ($s=\lceil n/4 \rceil$) dominates a naive approach without a hash-stride ($s=1$). This result appeared across all datasets, so for the remainder of the paper all results are shown with the hashing-stride of $s=\lceil n/4 \rceil$. In extended testing, we also investigated other ratios such as $s = n/2$ and $s=n$. While all $s=\bigO(n)$ performed better than $s=1$, the choice of $n/4$ seemed to consistently perform best among the options tested.  

\subsection{Computational Efficiency of KiloGrams}

Computing the top-$k$ most frequent $n$-grams has historically been computationally demanding, restraining most to consider only $n\leq 6$ unless working with small datasets. We have shown, from a theoretical view, that the KiloGram algorithm is $\bigO(L)$ complexity and practically fixed memory cost at $\bigO(B + k \cdot n)$. We now show that this result is matched empirically.

We measure runtime on a server with four Xeon E7-8870 CPUs for a total of 80 cores, 2 TB of RAM, and 40 TB of SSD storage. 
Because of the hashing-stride, we find that the runtime tends to decrease as $n$ increases. For the VS-20C corpus, computing 8-grams took 27 minutes whereas 1024-grams took only 12 minutes. 

While our primary results come from the use of a powerful server due to the need to train large logistic regression models, we note that such high-end equipment is not necessary to perform the $n$-gramming. The nature of the KiloGram algorithm means that any machine with $\approx$10 GB of RAM should have no difficulty in performing the computation. To emphasize this, we re-ran the same KiloGram code on a workstation with a 10 core Xeon E5-2650 at 2.30GHz, 128 GB of RAM, and a 4 TB SSD. It took only 41 minutes to compute
the 1024-grams on this machine.
Thee KiloGram algorithm can apparently run on modest hardware thanks to its computational and memory efficiency. 

Even if one is interested in small values of $n$, the KiloGram approach exhibits superior run-time complexity and can provide dramatic speedups over naive approaches. On the largest corpus,
Industry EXE dataset (2M files),  KiloGram took $\leq12$ hours of computation for all values of $n\le1024$. Mature code with three-years of performance tuning required \text{one month} to compute 6-grams in the classical way: a 60x speedup for Kilogram over this baseline.

\subsection{Investigating Large $n$ } 

As we discussed in \autoref{sec:related_work}, many have suggested the need for large $n$-gram sizes in building models for malware classification. However, after an extensive literature review, we found that no prior work  empirically evaluated large $n$-grams on a large modern dataset. We present the first evaluation of large $n$-grams, and show in \autoref{tbl:pred_performance} the balanced accuracy and AUC across all four datasets. The last "Ind-2-EMBER" columns show results applying a model trained on Industry EXE to the EMBER test set, making a strong test for durability against concept drift over three years. 

\begin{table}[ht]
\caption{Results as $n$ increases, using hashing-stride.}
\label{tbl:pred_performance}
\begin{adjustbox}{max size={1.00\columnwidth}{0.85\textheight}}
\begin{tabular}{@{}rccccccccc@{}}
\toprule
     & \multicolumn{2}{c}{Industry EXE} & \multicolumn{2}{c}{EMBER} & \multicolumn{2}{c}{Public PDF} & VS-20C  & \multicolumn{2}{c}{Ind-2-EMBER} \\ 
     \cmidrule(lr){2-3} \cmidrule(lr){4-5} \cmidrule(l){6-7} \cmidrule(l){8-8} \cmidrule(l){9-10}
$n$  & Acc             & AUC            & Acc         & AUC         & Acc            & AUC           & Acc     & Acc         & AUC       \\ \midrule
8    & 98.2            & 99.8           & 99.2        & 99.9        & 98.9           & 99.7          & 95.2    & 97.6        & 99.7       \\
12   & 97.5            & 99.7           & 98.9        & 99.9        & 98.7           & 99.7          & 93.8    & 97.4        & 99.4       \\
16   & 96.7            & 99.5           & 98.6        & 99.8        & 98.7           & 99.6          & 92.3    & 95.9        & 98.9       \\
24   & 96.4            & 99.4           & 97.9        & 99.7        & 98.6           & 99.6          & 88.1    & 95.5        & 98.4       \\
32   & 96.0            & 99.3           & 97.1        & 99.4        & 98.2           & 99.6          & 85.2    & 93.9        & 97.9       \\
64   & 94.9            & 99.1           & 96.3        & 99.2        & 92.0           & 99.3          & 87.4    & 92.9        & 96.8       \\
128  & 94.0            & 98.7           & 93.6        & 97.8        & 92.8           & 99.0          & 79.4    & 88.9        & 94.9       \\
256  & 92.6            & 98.0           & 90.3        & 95.6        & 91.3           & 98.5          & 76.5    & 86.6        & 91.9       \\
512  & 92.2            & 96.8           & 78.7        & 84.8        & 86.5           & 96.8          & 71.7    & 71.9        & 69.9       \\
1024 & 91.9            & 96.1           & 78.6        & 85.2        & 72.3           & 90.9          & 67.1    & 72.6        & 72.6       \\ \bottomrule
\end{tabular}
\end{adjustbox}
\end{table}

Across each dataset, we found that predictive accuracy does not increase beyond $n=8$. Indeed, the maximal performance on all metrics occurs at $n=8$. With some variation, we found that the performance in AUC degrades slowly for $n \leq 32$ across all datasets, but accuracy sometimes degrades faster. For example, the gap on the Public PDF dataset for $n=8$ and $n=32$ is 0.7 points, but is a more significant 10.0 points for the VS-20C corpus. 

More surprising was that 1024-grams had \textit{any} predictive utility at all, let alone reaching 90\%+ accuracy or AUC across many of our datasets. Our intuition was that $n$-grams for large $n$ would be extremely brittle, common accross only a few sample, and therefore ineffective for generalizing to new files. This was not the case, however, and suggests re-use (perhaps in the form of code, header information, resources, or compiler fingerprints) in EXE and PDF document formats that allow these $n$-grams to generalize. We also see from the "Ind-2-EMBER" experiment that these $n$-grams can generalize across \textit{years} of concept drift. At $n=8$, a small loss of 0.6 points occurs. As $n$ gets larger, the performance after three years drops faster, indicating that after $n\geq128$, they lose significant robustness to concept drift.

Given these results, we expect that as the size of $n$ increases, the features may begin to correspond to ever more specific indicators of benign or malicious intent. For example, use of the Windows API function "GetProcAddress" is a common indicator of maliciousness across many Windows PE malware samples and can be detected with $n\leq 6$ \cite{raff_ngram_2016}, but this indicator alone is not enough to detect malicious files since there are many benign use cases for this function. As $n$ becomes larger, we expect to see features that instead address sub-populations of malware, rather than the population at large.

\begin{figure}[!h]
\centering
\begin{tikzpicture}
\begin{axis}[
    xlabel=NNZ,
    ylabel=Balanced Accuracy,
    legend pos=outer south,
    legend columns=3,
    xmode=log,
    ]
    
    \addplot+[mark=none] table [skip first n=9, x=Average NNZ, y=exe.test.12gram.3stride.exact.jsat WeightedAccuracy, each nth point=10,col sep=comma] {exe_results/exe.12gram.3stride.exact.lasso.csv};
    
    \addplot+[mark=none] table [skip first n=9, x=Average NNZ, y=exe.test.16gram.4stride.exact.jsat WeightedAccuracy, each nth point=10,col sep=comma] {exe_results/exe.16gram.4stride.exact.lasso.csv};
    
    \addplot+[mark=none] table [skip first n=9, x=Average NNZ, y=exe.test.32gram.8stride.exact.jsat WeightedAccuracy, each nth point=10,col sep=comma] {exe_results/exe.32gram.8stride.exact.lasso.csv};
    
    \addplot+[mark=none] table [skip first n=9, x=Average NNZ, y=exe.test.64gram.16stride.exact.jsat WeightedAccuracy, each nth point=10,col sep=comma] {exe_results/exe.64gram.16stride.exact.lasso.csv};
    
    \addplot+[mark=none] table [skip first n=9, x=Average NNZ, y=exe.test.128gram.32stride.exact.jsat WeightedAccuracy, each nth point=10,col sep=comma] {exe_results/exe.128gram.32stride.exact.lasso.csv};
    
    \addplot+[mark=none] table [skip first n=9, x=Average NNZ, y=exe.test.256gram.64stride.exact.jsat WeightedAccuracy, each nth point=10,col sep=comma] {exe_results/exe.256gram.64stride.exact.lasso.csv};
    
    \addplot+[mark=none] table [skip first n=9, x=Average NNZ, y=exe.test.512gram.128stride.exact.jsat WeightedAccuracy, each nth point=10,col sep=comma] {exe_results/exe.512gram.128stride.exact.lasso.csv};
    
    \addplot+[mark=none] table [skip first n=9, x=Average NNZ, y=exe.test.1024gram.256stride.exact.jsat WeightedAccuracy, each nth point=10,col sep=comma] {exe_results/exe.1024gram.256stride.exact.lasso.csv};

    \legend{12-grams, 16-grams, 32-grams, 64-grams, 128-grams, 256-grams, 512-grams, 1024-grams} %
    
\end{axis}
\end{tikzpicture}
\caption{Balanced Accuracy results (y-axis) on the Industry EXE dataset as the number of non-zero weights (x-axis) learned by the logistic regression model increases.}
\label{fig:industry_exe_results}
\end{figure}
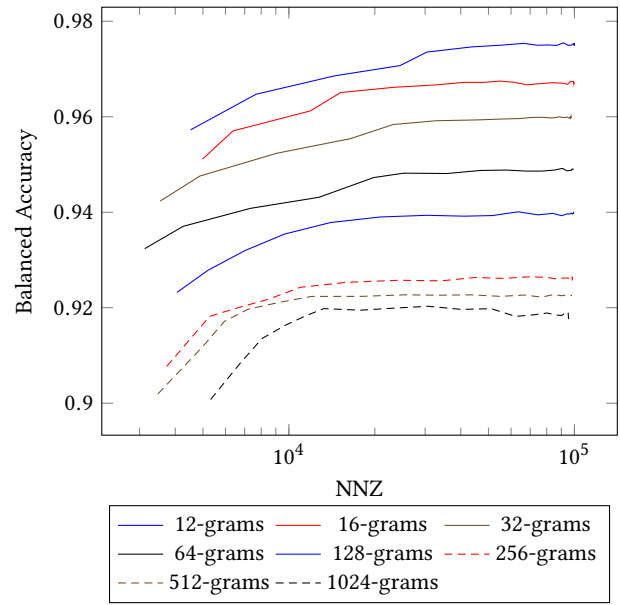

In \autoref{fig:industry_exe_results}, we plot the balanced accuracy as a function of the number of non-zero (NNZ) weights in the learned Elastic-Net regularized logistic regression model, where fewer non-zero features corresponds to a larger regularization penalty $\lambda$. Here we see that for small $n$, there is a smooth and continuous increase in accuracy as more features are selected along the regularization path.  As $n$ increases, the behavior transitions to an initial rise in accuracy, followed by a plateau once a minimum number of features are obtained. The start of this plateau occurs earlier and the initial slope larger as $n$ becomes larger. 

This behavior is intuitive, and corresponds with our expectation that the specificity of $n$-grams will increase with their size. For small values, a large number of $n$-grams are necessary to cover a wide range of smaller components that reflect the work and actions of larger features. At larger values of $n$, the model quickly selects all "useful" features. We believe this is because it becomes easier to delineate the predictive subset due to feature occurrences clustering around increasingly specific subsets of the population. Once a sub-population is well separated, additional features have little value unless they can "carve off" a different sub-population, and so performance plateaus. A unique benefit of larger $n$-grams is their increased interpretability, which allows us to provide additional evidence to  this interpretation of our results in \autoref{sec:analysis}. 

\section{Feature Analysis} \label{sec:analysis}
The previous sections described how KiloGrams are computed, and how they perform as features in a machine learning algorithm. For malware classification we find large KiloGrams have considerably more value in their application to analyst work-flow and integration into larger systems.
In this section we will describe how malware analysts can use larger $n$-gram features in the course of their investigations, how they can be used in current signature based tools like Yara, and how they can be integrated with domain knowledge features to build a 
more powerful malware classification
system. In going through a number of $n$-gram features, both experienced and junior analysts determined that it usually took a few minutes to understand what a single feature meant or represented, with some features taking longer. 

\subsection{Analyzing Individual Features} \label{sec:indiv}
In a machine learning context, there are many features that could be pulled from binary files for use in classification, such as information from the PE header, printable strings, or (in our case) raw byte sequences. Malware analysts, whose job usually involves dissecting pieces of malware to write detection signatures or understand how they operate, work with many of these feature types in the course of these investigations. From the point of view of these analysts, a machine learning system based on small $n$-grams is opaque; it takes many of these features to compile enough evidence to apply a benign or malicious label, and since these $n$-grams may be only a portion of a single x86 instruction as discussed earlier, they are not very helpful in shedding light on why the algorithm chose the label it did. When working with malware analysts in production, the inability to understand what a particular feature means when it is found in a binary has been a source of frustration, and impeded adoption. 

The sheer size of KiloGrams changes this dynamic and presents a way to interpret features in what may otherwise be a black box. The presence of a KiloGram provides an immediate indication of where to start looking, and a large enough section of bytes to be meaningful to an analyst.
Features that contribute the most to a malicious/benign decision can contain strings, embedded data like images, or code (that may or may not require disassembly or decompilation\footnote{Disassembly is the process of converting raw bytes into the low-level assembly language they represent, while decompilation converts assembly instructions into a higher-level language like C.}, depending on file format). This can help analysts reach a conclusion about a binary's nature, which is important since the average time to process a malicious file is 10 hours \cite{Mohaisen:2013:UZA:2487788.2488056}. The ability to understand what the features mean is also important to build trust so that developed solutions will be adopted and used. Below we will provide some examples of the interpretability of features found by malware analysts at different $n$-gram sizes. 

\subsubsection{EMBER Examples with 64-grams}\label{sec:ember}

In \autoref{fig:asm_64gram_example}, we see the disassembly of a code snippet discovered by a 64-gram that occurred in 8\% of the EMBER dataset. Upon inspection this code assembles the string "VirtualAlloc". This is then later used to obtain the "GetProcAddr" function in an obfuscated manner, so that the binary can then load other libraries at runtime. This is a technique to obfuscate the true intentions of the binary from malware analysts, and is considered a strong indicator of maliciousness. \todo{cite PMA?}

\begin{figure}[!ht]
    \centering
\begin{minted}[fontsize=\footnotesize,frame=lines,framerule=1pt,breaklines]{gas}
push eax #50
call DWORD PTR [ebp-0xcc] #ff9534ffffff
mov DWORD PTR [ebp-0x20], eax #8945e0
mov DWORD PTR [ebp-0xa4], 0x74726956 #c7855cffffff56697274
mov DWORD PTR [ebp-0xa0], 0x416c617f #c78560ffffff75616c41
mov DWORD PTR [ebp-0x9c], 0x636f6c6c #c78564ffffff6c6c6f63
and DWORD PTR [ebp-0x98], 0x #83a568ffffff00
lea eax, [ebp-0xa4] #8d855cffffff
push eax #50
push DWORD PTR [ebp+0xe] #ff750e
xor bh,bh #30ff
xchg ebp,eax #95
cmp bh,bh #95
.byte 0xff #ff
\end{minted}
    \caption{Example of a disassembled 64-gram feature found in the EMBER dataset. The hex values of the raw bytes are shown in comments for each line of assembly. }
    \label{fig:asm_64gram_example}
\end{figure}

A considerable number of 64-grams contained sub-strings of registry keys. In the EMBER dataset, 10\% of malware was found to have
\texttt{\seqsplit{HKEY\_LOCAL\_MACHINE\textbackslash{}SOFTWARE\textbackslash{}Microsoft\textbackslash{}Windows\textbackslash{}CurrentVersion\textbackslash{}Policies\textbackslash{}Explorer}}, which may be used to hijack the explorer process. Another 12\% contained \texttt{\seqsplit{HKEY\_LOCAL\_MACHINE\textbackslash{}Software\textbackslash{}Microsoft\textbackslash{}Windows\textbackslash{}CurrentVersion\textbackslash{}Run}} for securing persistence for the malware. 

One 64-gram triggered on the image icon shown in \autoref{fig:malware_icon} in \autoref{sec:additional_figures} of the Appendix. 
Several malware samples contained precisely this icon in the resources section, causing it to become a predictive feature. This is another example of a feature with an easily understood nature,  thanks to the large scale of 64-grams, which would have been uninformative if broken up into standard $n\leq 6$ grams.

\subsubsection{PDF Example from 512-grams}\label{sec:pdf512}
For an example of a 512-gram, \autoref{fig:obf_javascript} is a fragment from our Public PDF dataset discovered by simply looking through the extracted features. 
The scripting code was identified as JavaScript, which is often used in PDF files, but more frequently used (and obfuscated, as in this case) in malicious PDF files. 
The long string in the middle also stood out; 0x41 is the letter 'A' in hexadecimal format, and long strings of this character are often used by exploit writers to assist in crafting the correct string to take advantage of a vulnerability.

\newminted{javascript}{fontsize=\footnotesize,frame=lines,framerule=1pt,breaklines,breakanywhere}
 
 \begin{figure}[!ht]
    \centering
\begin{javascriptcode*}{}
ey=aba();ek=24;ek++;ef=6192;ef--;if(ey>=wf){ep=4810;if(ep!=null){ev=3742;ev+=0.013}eq=0.03;eq--;ed=we;ea=0.017;if(ea==0.0161){er='tae'}ex=sub(wd,wu);ez=5544;ez++;ej=true;es=wy;eu=8;eu--;eb='';en=9;en++;eh='';eo=0.034;eo++;ei=6343;ei++; if(ey&lt;f){el=8952;el--;eb=r;mt=6172;mt+=0.0101;mw=1834;mw-=7115;eh='4c20600f0517804a3c20600f0f63804aa3eb804a3020824a6e2f804a41414141260000000000000000000000000000001239804a6420600f000400004141414141414141';me=6;me++}else if(ey&lt;h){mm=18;mm++;eb=u;mg=false;mc=[7,35,21,4wp>null){wv=0.0082;wv++}wf+=2177;wq=0.0032;wq+=3755;wd=zoa('qXM7reN6',15);wa=0.006;if(wa!=21){wr=0.014;wr++}wx=zoa('tCTF3OdREync',13);wz=[40,24,32,8,48,0,16,56];wj=0.008;if(wj&lt;0){ws=null;ws+=12}wu=18690;wb=0.007;if(wb!=13){wn=[0,16,8,24];wh=17;wh+=7832}wu-=7706;wo=28;if(wo==8){wi=['fen','lag','het']}wl=22;if(wl&lt;4370)
\end{javascriptcode*}
    \caption{
    A 512-gram found in PDF dataset. This example contains obfuscated JavaScript code to build an exploit string targeting particular versions of PDF reader software. }
    \label{fig:obf_javascript}
\end{figure}

The next step in analysis is to find and extract the entire piece of embedded code.  In this case a web search yielded the exploit code in a public repository. The code was then manually deobfuscated which revealed its functionality. Two different versions of an exploit string are built, depending on the software version (the target is likely to be Adobe Acrobat, the most popular PDF reader, but this was not confirmed). A successful exploit results in a visit to http://phjqxagpgdw.com/nte/goldmn.asp; and a quick search revealed a number of identified malicious domains with this naming scheme.

\subsection{Features as Signatures} \label{sec:yara}

As the value of $n$ increases, the resulting features gain increased specificity in the files they target. It also becomes less likely to observe an exact $n$-byte sequence by chance. This inspired us to explore how well some of these larger $n$-grams might serve as Yara signatures for malware detection. Yara \cite{Alvarez2013} is an industry standard regular expression tool designed for malware analysis. Yara rules are usually designed to have low false-positive rates, in the sense that if a specimen causes a rule to match, the specimen is probably malicious. Using KiloGrams this way may seem counter-intuitive: Our $n$-grams make powerful features for machine learning algorithms, which many believe will ultimately replace signature-based malware detectors. However, signature-based systems are still widely used, and will likely have a large role to play in layered defensive systems for the foreseeable future.

YarGen\cite{Roth2013} is the only currently-maintained tool we know of for automatically generating Yara signatures for Windows PE files.  YarGen uses a number of domain knowledge processing steps to create a signature from several files. Where it is feasible, we use YarGen as a comparison to our method. \todo{lonely paragraph? - better? wrf}

For our approach, we will use the coefficients learned by logistic regression to select the 4000 $n$-grams most indicative of the class we are interested in. We then look at the false positive rate of each individual $n$-gram (on the training data), and discard any with a false-positive rate above 5\%. If any two or more $n$-grams \textit{always} co-occur, we randomly select one of the $n$-grams and discard the others. We then combine the remaining set to form a simple Yara rule, which looks for the exact $n$-grams, and fires if any of the $n$-grams occur. Normally a combination of sub-rules is necessary; for example, YarGen usually only fires if 3 or more out of a list of patterns match.
\todo{cannot parse - better? wrf}
However, the statistical improbability of any individual KiloGram makes them independently robust detectors. After creating a rule for many values of $n$, we select the rule with the best $F_1$ score on the training set.

\subsubsection{Results on VirusShare-20C Dataset}\label{sec:yara_family}

\begin{figure}[]
\centering
\begin{adjustbox}{max size={\columnwidth}{\textheight}}
  \begin{tikzpicture}
  
	\begin{axis}[
                ybar,
                /pgf/bar width=2pt,
                ylabel=$F_1$ Score,
                symbolic x coords={azero,bancos,buzus,dapato,domaiq,hoax,inbox,karagany,nebuler,pasta,pcclient,scar,shipup,sillyfdc,softpulse,spyeye,toggle,vbinder,zegost,zwangi},
                xtick=data,
                xticklabels={azero,bancos,buzus,dapato,domaiq,hoax,inbox,karagany,nebuler,pasta,pcclient,scar,shipup,sillyfdc,softpulse,spyeye,toggle,vbinder,zegost,zwangi},
                x tick label style={rotate=45, anchor=east, align=right},
                legend style={at={(1,1.15)},anchor=north east},
                legend columns=2,
                ymin=0,
                ymax=1.15,
                enlarge x limits={0.05},%
                x tick label style={font=\small} ,
                height=4cm,
                width=\columnwidth,
            ]

            \addplot table[x=family,y=yg_f1,col sep=comma]{other_results/yara_family_all_results.csv};
			\addplot table[x=family,y=kilo_f1,col sep=comma]{other_results/yara_family_all_results.csv};

            \legend{YarGen-$F_1$,KiloGram-$F_1$}

		\end{axis}
\end{tikzpicture}
\end{adjustbox}
\caption{The $F_1$ score of Yara rules automatically generated from the VirusShare-20C training data, and evaluated on the test data. A different rule set is created for each family. }
\label{fig:yara_family_results}
\end{figure}
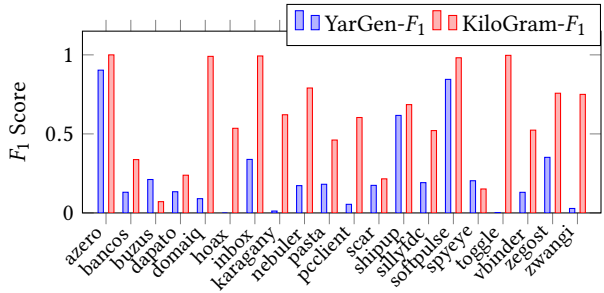

First we used the malware family dataset described in \autoref{sec:results} to automatically create Yara rules to identify specific families. This is a common and arduous task normally done manually by a malware analyst. 
We trained one family-vs-the rest for $n \in [8, 1024]$, and found that no single value of $n$ was best for all families. For 9 out of 20 families, $n=1024$ did perform best, which is a trend counter to the use of KiloGrams as purely predictive features in \autoref{sec:results}. 
We compared the results of our new approach to the existing YarGen in \autoref{fig:yara_family_results}, where we look at the $F_1$ score for each family. A Wilcoxon signed rank test \cite{Demsar:2006:SCC:1248547.1248548} shows that our KiloGram based approach is better, with p-value  of $3.1 \times 10^{-5}$.  %

These results should not be taken to mean that YarGen is inferior to KiloGrams; YarGen is a tool that iterates over various features (strings, byte opcodes, etc) to find the best predictive rules. In a sense, KiloGrams represent a new class of features that tools such as YarGen could incorporate to improve their detection rates. We believe these results show that KiloGrams can be a valuable tool in the creation of signatures for malware detection, and the combination of these large features with machine learning tools can help automate the process of signature creation. 

For a final test the Yara rules used for each family were run over the EMBER benign test set, as having low false positives on benign files is a critical feature. Note that no benign datasets were used in the creation of these KiloGrams.
The KiloGram based rules had a median false-positive rate of 0.0065\%. This is $24\times$ better than YarGen, which had a median false-positive rate of 0.1595\%.

\subsubsection{Results on Industry EXE Dataset.}\label{sec:yara_industry}
We repeated the same experiments on our Industry EXE dataset, to create a \textit{generic} "malware" signature. This is unprecedented in the standard use of Yara, which is meant for identifying files of a \textit{specific} nature. YarGen failed to run on 2 million files in a timely fashion, so we are unable to compare with any prior works in the goal of creating a generic malware signature. The results when attempting to use different values of $n$ are shown in \autoref{tbl:yara_conf}. 

\begin{table}[ht]
\caption{Yara generation results on Industry EXE}
\label{tbl:yara_conf}
\begin{adjustbox}{max size={1.00\columnwidth}{0.85\textheight}}

\begin{tabular}{@{}rccccccc@{}}
\toprule

n                   &\# of rules   & True Neg    & False Pos    & False Neg    & True Pos    & Precision    & Recall       \\ \midrule
32                  & 25           & 36.643\%    & 63.357\%     & 15.742\%     & 84.259\%    & 57.08\%       & 84.26\%       \\
64                  & 23           & 81.554\%    & 18.446\%     & 56.791\%     & 43.209\%    & 70.08\%       & 43.21\%       \\
128                 & 22           & 98.599\%    & 1.401\%      & 78.44\%      & 21.56\%     & 93.90\%       & 21.56\%       \\
256                 & 4            & 100\%       & 0\%          & 95.034\%     & 4.966\%     & 100.00\%      & 4.97\%        \\
512                 & 31           & 99.987\%    & 0.013\%      & 78.777\%     & 21.223\%    & 99.94\%       & 21.22\%       \\
1024                & 52           & 99.947\%    & 0.054\%      & 76.68\%      & 23.32\%     & 99.77\%       & 23.32\%       \\
2048                & 35           & 99.989\%    & 0.012\%      & 78.392\%     & 21.609\%    & 99.95\%       & 21.61\%       \\
4096                & 84           & 99.992\%    & 0.009\%      & 89.79\%      & 10.21\%     & 99.92\%       & 10.21\%       \\
8192                & 145          & 99.684\%    & 0.316\%      & 89.61\%      & 10.39\%     & 97.05\%       & 10.39\%       \\
$[$256-4096$]$            & 206          & 99.938\%    & 0.063\%      & 74.958\%     & 25.042\%    & 99.75\%       & 25.04\%       \\
\bottomrule
\end{tabular}
\end{adjustbox}
\end{table}

We can see that any $n\in [256, 4096]$ produces signatures with low false positive rates, and surprisingly can catch up to 23\% of the malware in the test set. Naively combining all KiloGrams in this range into one larger signature of $256$ through $4096$-grams boosts the recall up to 1/4 of the test set malware. This produced 125 false positives, which we investigated with VirusTotal\cite{Virustotal}.
Of these, 10 are now reported as \texttt{OutBrowse} malware; 45 behave very similarly to each other and are almost certainly adware/spyware; 10 are unsigned (a huge red flag) versions of mmc.exe (Microsoft Management Console) in various languages; and another 11 have other malicious indicators (such as 1 or more malicious AV reports or relationships with other malicious files). Only 49 of the 125 reported false positives display no evidence of being malware. 

Further, we tested the Industry EXE generated signatures on the EMBER test set, which is 2-3 years newer. This signature was still able to catch 8.7\% of the EMBER malware, with a false positive rate of 0.0093\% on the EMBER benign set. 
The ability of these signatures to catch mislabeled data in our test set, and still generalize to data three years later (despite the concept drift common in this domain) increase our confidence in the usefulness of KiloGrams as signatures. 

\subsection{KiloGrams \& Domain Knowledge} \label{sec:kilo_domain}

Based on the analysis in \autoref{sec:indiv} we find large $n$-grams represent interesting and relevant features present in large sub-populations of the malicious or benign binaries. This leads us to ask, can combing large $n$-gram features with human-engineered features produce a stronger model? 

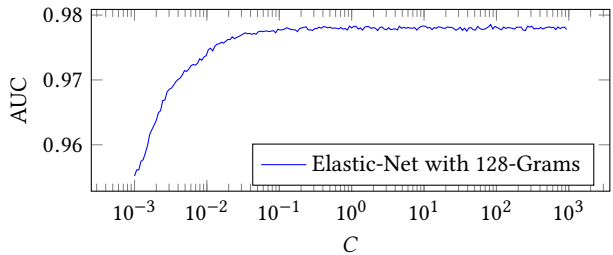
\begin{figure}[!h]
\begin{center}
\begin{tikzpicture}
\begin{axis}[
    xlabel=$C$,
    ylabel=AUC,
    legend pos=south east,
    legend columns=1,
    xmode=log,
    height=4cm,
    width=\columnwidth,
    ]
    
    \addplot+[mark=none] table [x=C, y=AUC,,col sep=comma] {featurecompare_results/kilogram_ember_selection_t.csv};

    \legend{Elastic-Net with 128-Grams} %
    
\end{axis}
\end{tikzpicture}
\end{center}
\caption{
AUC
as a function of regularization parameter $C$ using Elastic-Net to force most coefficients to zero. Selecting $C=10^{-1}$ gave 17,294 nonzero 128-gram features.}
\label{fig:featureselect}
\end{figure}

\begin{figure}[!t]
\begin{center}
\centering
\begin{tikzpicture}[spy using outlines=
	{rectangle, magnification=8, connect spies}]
\begin{axis}[
    xlabel=False Positive Rate,
    ylabel=True Positive Rate,
    legend pos=south east,
    legend style={line width=2.5pt,draw=none},%
    legend columns=1,
    xmode=log,
    xmax=0.1,
    height=6cm,
    width=\columnwidth,
    ]
    
    \addplot+[Brown,mark=none,line width=2.0pt] table [x=KiloGram FPR, y=KiloGram TPR, each nth point=10,col sep=comma] {featurecompare_results/kilogram_industry_comparison_roc_t.csv};
    
    \addplot+[LimeGreen,mark=none] table [x=EMBER FPR, y=EMBER TPR, each nth point=10,col sep=comma] {featurecompare_results/kilogram_industry_comparison_roc_t.csv};
    
    \addplot+[OliveGreen,mark=none] table [x=EMBER+KiloGram FPR, y=EMBER+KiloGram TPR, each nth point=10,col sep=comma] {featurecompare_results/kilogram_industry_comparison_roc_t.csv};
    
    \addplot+[Dandelion,mark=none] table [x=Industry FPR, y=Industry TPR, each nth point=10,col sep=comma] {featurecompare_results/kilogram_industry_comparison_roc_t.csv};
    
    \addplot+[Red,mark=none] table [x=Industry+KiloGram FPR, y=Industry+KiloGram TPR, each nth point=10,col sep=comma] {featurecompare_results/kilogram_industry_comparison_roc_t.csv};

    \legend{128-grams, EMBER, EMBER+128-grams,Proprietary,Proprietary+128-grams} %
    
    \coordinate (spypoint) at (axis cs:0.0005,0.98);
    \coordinate (magnifyglass) at (axis cs:0.0004,0.8);
    
\end{axis}

\spy [black,width=2.2cm,height=1cm] on (spypoint)
   in node[fill=white] at (magnifyglass);

\end{tikzpicture}
\end{center}
\caption{KiloGrams augment the EMBER features to create a model that rivals one built using proprietary features.}
\label{fig:featurecompare}
\end{figure}
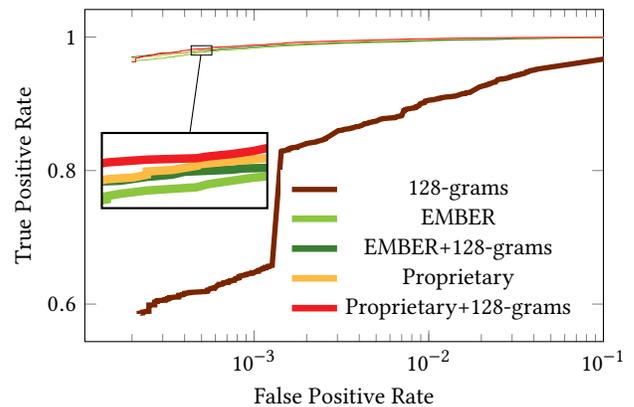

To assess this, the top 100,000 128-grams were extracted from the EMBER training files.  
Using the Elastic-Net regularization path on the training data 
(\autoref{fig:featureselect}),  %
we selected the regularization parameter, $C$, corresponding to the sparse subset that maximizes AUC.  This resulted in $C=10^{-1}$ and 17,294 nonzero 128-gram features.  We measured the lift provided by these 128-gram features by prepending the counts to two different domain knowledge feature sets: 
the 2,351 EMBER features crafted via domain knowledge, as well as a production set of feature extractors designed by malware analysis experts.  These were compared to 128-grams alone, EMBER features alone, and to proprietary features alone.
We trained a gradient-boosted decision tree model using \texttt{xgboost} on each of these feature sets, with 200 boosting rounds, tree depths up to 9 levels, 50\% column subsampling per tree, and a $\eta=0.29$ learning rate \cite{xgboost}. ROC curves on the validation features are shown in \autoref{fig:featurecompare}.

Adding 128-grams improved AUC in all cases. EMBER features alone achieved an AUC of 0.999597, and improved to 0.999718 when augmented with 128-grams. 
The proprietary features result in a slightly higher AUC at 0.999822, and further improved to 0.99985 when augmented with 128-grams. 

For a production malware detector deployed as an anti-virus, we care about the true positive (TPR) rate at n very low false-positive rate (FPR). The zoomed inset in \autoref{fig:featurecompare} shows the TPR at at FPR of 5:10000, which is reasonable for a production system. At that rate, the TPR of EMBER with 128-grams is comparable to the proprietary features alone, and then further outperformed by proprietary features with prepended 128-gram counts. Also of interest is the ROC curve for 128-grams alone, which exhibits a peculiar jump in TPR near $2\times10^{-3}$ FPR.  This fits intuition that KiloGrams are essentially ``getting the easy ones'' via the top $k$ $n$-grams spanning large subsets of malicious or benign PE files.  Feature combinations involving domain knowledge features cover the remaining samples.  Indeed, it required 20 boosting rounds for a model trained on EMBER features to exceed 0.999 AUC on the evaluation set, but only 15 boosting rounds when augmented with KiloGrams.

While the EMBER dataset has been noted as a relatively "easy" dataset \cite{Anderson2018}, the results are a positive indicator of the utility of large $n$-grams in conjunction with domain knowledge features. Based on these promising results, more extensive work is being prepared to test KiloGram augmented production features on industry-representative datasets.

\section{Conclusion} \label{sec:conclusion}

We have introduced the KiloGram algorithm for computing the top-$k$ most frequent $n$-grams. It is over 60x faster than prior approaches for small $n$, and allows the new capability to select $n\geq 1024$ with no increase in runtime. This allows us to explore previously unanswerable questions about large $n$-grams for malware classification. More careful consideration about the nature of such large $n$-grams allows us to address several issues in real-life malware analysis work. We can create new features that are interpretable and increase analyst trust, automate the creation of signatures with greater precision and recall than was previously possible, and enhance the detection rate of production anti-virus malware detectors.

Our belief is that the introduced ability to use $n\geq16$ grams will lead to interesting new ways of using $n$-grams, and may be applicable in natural language processing, network traffic analysis, bioinformatics, and other fields. 
For malware detection, 
malware authors trying to evade detection can no longer just obfuscate strings or header fields. They must now consider any potential binary pattern, such as compiler tags or code reuse, increasing the effort for them to operate undetected.

\bibliographystyle{ACM-Reference-Format}
\bibliography{Mendeley,NicholasMendeley}

\newpage
\begin{appendices}

\section{Supplementary Material}

\subsection{Dataset Source Details}

For completeness and reproducibility, we will review the nature of our datasets in greater detail. 

The "Industry EXE" dataset was provided by a third party AV company. The training set contains approximately 2 million Windows PE executables, 1,000,020 benign and 1,011,766 malicious, and was used as the training data in \cite{raff_shwel}. A smaller corpus of 400,000 files from the same AV company  as a large test set, which is also evenly split and was used as the training data in\cite{raff_ngram_2016}. All the executable binaries in this corpus were first seen in the 2014-2015 time frame. This company has asked to remain anonymous, but allowed the use of their older data. 

The EMBER corpus \cite{Anderson2018} normally provides the SHA256 hashes, VirusTotal outputs, and domain knowledge features pre-extracted for the public. The raw files of EMBER were obtained from VirusTotal\cite{Virustotal}. This corpus has 600k training and 200k testing files which are evenly split between benign and malicious. The training corpus files contain a small subset from 2016, but the majority (and all test files) are from 2017. The entirety of the EMBER test set has a first-observed date newer than anything in the training data to make a better test of generalization. As mentioned previously, the whole corpus is on average 2-3 years newer than the Industry EXE dataset. It was also collected and organized by a different company, with no collaboration. Using it as a cross-dataset generalization test is then particularly powerful and informative to the longevity of extracted features and models, as we minimize common source bias and generalization occurs after at least 2 years of separation. This is the longest scale test of generalization across time we are aware of in this domain. 

Our PDF dataset was constructed from publicly available sources. As such, this corpus may not represent benign and malicious PDF populations in the same way as data collected by AV companies. This is because the AV companies can observe a subset of real-life benign and malicious traffic as they occur on real networks, where our collection from public resources may reflect different sub-populations.  

For the malicious PDFs, we downloaded the VirusShare corpus of malware \cite{VirusShare}, and selected all files in the corpus that were PDFs. The VirusShare dataset is mostly Windows PE data, and so we are only able to collect a total of 157,780 malicious files. For our benign files, we used Common Crawl\footnote{\url{http://commoncrawl.org}}, a non-profit effort to produce a publicly available "crawl" of the internet similar to that used by search engines. We randomly downloaded PDF files indexed in the common crawl to create our benign files. The entire corpus is about 19\% malware. We are aware that assuming all files indexed by common crawl are benign may not be absolutely true, but several spot checks did not turn up any obviously malicious files. The inspection of KiloGrams from \autoref{sec:analysis} also gives us confidence that this is not a systematic problem, as we would otherwise be unlikely to learn such useful and interpretable features from the PDF corpus if contamination did occur. 

Our last dataset, VirusShare-20C, is a malware family classification  dataset so that we could study a multi-class problem instead of a binary one. This dataset was constructed by again using the public VirusShare corpus \cite{VirusShare}. To determine the malware families, we use the VirusTotal \cite{Virustotal} Anti-Virus labels provided by  \cite{Seymour2016}. The VirusTotal results include a label from several different Anti-Virus products. Each AV product may use different naming schemes, have conflicts, and sometimes different sets of AVs run against each file. We use AVclass\cite{Sebastian2016} to take the results from VirusTotal and produce a single canonical family name and label for each file. This produced 184 families which each had at least 10,000 samples. 20 of these were selected at random to create our dataset. For each family, 8,000 files were used for the training set and 2,000 for the testing set.

\subsection{Additional Figures}
\label{sec:additional_figures}

\begin{figure}[!h]
\begin{center}
\includegraphics[width=0.25\columnwidth]{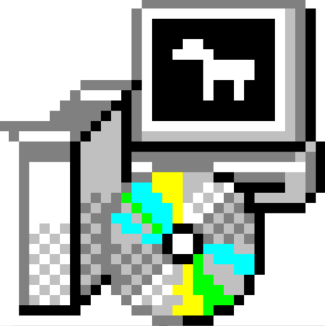}
\end{center}
\caption{Frequently-used malware icon, found by inspecting 64-gram features of Lasso.}
\label{fig:malware_icon}
\end{figure}

\end{appendices}

\end{document}